\documentclass[conference]{IEEEtran}
\usepackage[utf8]{inputenc}
\usepackage[frozencache,cachedir=.]{minted}

\IEEEtriggeratref{21}

\usepackage{cite}
\usepackage{amsmath,amssymb,amsfonts}
\usepackage{algorithmic}
\usepackage{graphicx}
\usepackage{textcomp}
\usepackage{xcolor}
\usepackage{url}
\usepackage[normalem]{ulem}

\usepackage{boxedminipage}
\newenvironment{result}
{\smallskip
\noindent
\let\emph=\textbf
\begin{boxedminipage}{\columnwidth}\em}
{\end{boxedminipage}
}

\begin{document}

\title{Automatic Property-based Testing of GraphQL APIs}

\author{\IEEEauthorblockN{Stefan Karlsson}
\IEEEauthorblockA{ABB AB, Mälardalen University\\
Västerås, Sweden\\
Email: stefan.l.karlsson@\{se.abb.com, mdh.se\}}
\and
\IEEEauthorblockN{Adnan \v{C}au\v{s}evi\'{c}}
\IEEEauthorblockA{ABB AB, Mälardalen University\\
Västerås, Sweden\\
Email: adnan.causevic1@se.abb.com}
\and
\IEEEauthorblockN{Daniel Sundmark}
\IEEEauthorblockA{Mälardalen University\\
Västerås, Sweden\\
Email: daniel.sundmark@mdh.se}}

\maketitle

\begin{abstract}
In recent years, GraphQL has become a popular way to expose web APIs. With its raise of adoption in industry, the quality of GraphQL APIs must be also assessed, as with any part of a software system, and preferably in an automated manner. However, there is currently a lack of methods to automatically generate tests to exercise GraphQL APIs.

In this paper, we propose a method for automatically producing GraphQL queries to test GraphQL APIs. This is achieved using a property-based approach to create a generator for queries based on the GraphQL schema of the system under test.

Our evaluation on a real world software system shows that this approach is both effective, in terms of finding real bugs, and efficient, as a complete schema can be covered in seconds. In addition, we evaluate the fault finding capability of the method when seeding known faults. 73\% of the seeded faults where found, with room for improvements with regards to domain specific behavior, a common oracle challenge in automatic test generation.

\end{abstract}{}


\maketitle

\section{Introduction}

GraphQL\footnote{https://graphql.org/} is a technology publicly introduced by Facebook in 2015 \cite{GraphQLFoundation}. GraphQL provides a method to state declarative, compositional and strongly typed queries for Web APIs. Since its inception, GraphQL has become a popularly used alternative to RESTful APIs. Several large internet services provide parts of their APIs via GraphQL \cite{WhoIsUsingGraphQL}. Notable examples are GitHub\footnote{https://github.com/}, GitLab\footnote{https://about.gitlab.com/}, Shopify\footnote{https://www.shopify.com/}, Facebook\footnote{https://www.facebook.com/} and Pinterest\footnote{https://www.pinterest.com/}. Industry adopters, such as Shopify and GitHub, express benefits of moving to GraphQL \cite{shopifyGQLBenefits, GitHubGQLBenefits} and some of those benefits are supported in the literature. For example, Brito et al. report that in their experiments the returned payload size was reduced by up to 99\% by porting several RESTful client calls to GraphQL \cite{BritoMigratingtoGraphQL}. In addition, in a controlled experiment subjects spent less time on client implementation using GraphQL versus REST \cite{Brito-RESTvsGraphQL-2020}. Some industry adopters, GitLab for example, states that GraphQL will be their primary interaction model for their APIs going forward \cite{GitLabGraphQL}.

\begin{figure}[h]
    \centering
\begin{minted}{text}
{ person { pet { name } } }
\end{minted}
    \caption{Query example in GraphQL}
    \label{fig:person-gql-example}
\end{figure}

\begin{figure}[h]
    \centering
\begin{minted}{json}
{"data": {
   "person": {
     "pet" : {    
       "name": "Rex"}}}}
\end{minted}
    \caption{Example query result}
    \label{fig:person-gql-example-payload}
\end{figure}

RESTful APIs \cite{fielding2000} have been, and still are, a common approach to expose APIs of internal and external software systems. As a result of this, methods to automatically test RESTful APIs have been proposed. Some recent examples of methods that automatically produce tests, targeting RESTful APIs, are EvoMaster~\cite{Arcuri:2019:RAA:3292526.3293455}, RESTler~\cite{Atlidakis:2019:RSR:3339505.3339600}, QuickREST~\cite{karlsson2019quickrest} and RESTTESTGEN\cite{RESTTESTGEN}. However, with the increased usage to expose Web APIs using GraphQL, there are new challenges in testing Web APIs, which calls for new testing methods in addition to Web APIs using REST.

GraphQL and REST expose Web APIs in fundamentally different ways, one provide the clients with a query language and the other operations on resources. Other than exposing Web APIs commonly using JSON as format, there is not much in common between the two.
GraphQL enables a client to query only the data it requires. The client is provided with a typed schema of the possible entities to query and their fields. Figure \ref{fig:person-gql-example} shows an example query expressed in GraphQL and Figure \ref{fig:person-gql-example-payload} shows the result from executing such a query. Note how only the data explicitly asked for is returned and the layout of the result corresponds to the client's query. This is in contrast to REST, where querying the same information might require several operations.

Given the benefits of GraphQL and its rise in popularity in industry, it is of high interest to find methods to automatically produce test cases for GraphQL APIs, as has been done for RESTful APIs. Such methods must be able to produce GraphQL queries, conforming to the schema, covering the complete schema under test and handle the graph nature of the schema. 

In this paper, we introduce a method for automated testing of GraphQL APIs. We leverage property-based testing techniques to randomly generate test cases in the form of GraphQL queries. The generated queries are then executed on the API under test and the result is verified to not have caused any server error and to conform to the schema. In addition, we propose a method of how to measure the coverage of a GraphQL schema achieved by test cases. In this paper we contribute, to the best of our knowledge, with the first proposed method of how to fully automatically produce test cases, and measure schema coverage, for GraphQL APIs. We also present the findings of evaluating the method on GitLab, a large open-source software system, where we found and reported several new bugs. We have evaluated both the fault finding capability of the method as well as the coverage of the schema for the produced test cases.

\section{Background}
In this section, we present an overview of the GraphQL specification\footnote{https://spec.graphql.org/June2018/} and of property-based testing. For GraphQL, we restrict ourselves to the parts of the specification that are of relevance to understanding this paper and refer the interested reader to the full specification for a more in-depth view.

\subsection{GraphQL}

GraphQL is a query language for APIs. It was internally developed by Facebook, starting in 2012, and was open-sourced in 2016. Today, GraphQL is curated by the GraphQL Foundation who aims at increasing its adoption \cite{GraphQLFoundation}. GraphQL provides a language that allows clients to query an API for only the data needed by the client. This is both more efficient and more predictable for the client since the client knows the structure of the returned query result. This property both reduces the size of the result and the number of network calls. It also lessens the complexity of the client implementation, since the client does not need to keep track of intermediate query results, as is common in a RESTful API.

GraphQL is, as its name hints to, graph-based. This means that a query follows references of queried resources and a client can thus query a graph of related resources. To enable clients to inspect what resources are available to query, GraphQL has a type system that is accessible to clients. Clients can use query introspection to access meta-data about the queries, types, and fields available. Figure \ref{fig:schema-example} shows an example result of an introspection query that has queried for the available types in the schema. The result includes the meta-data information of the \texttt{Project} type and its fields.

A query in GraphQL defines a set of \textit{fields} of \textit{objects} to be queried. In the example of Figure \ref{fig:person-gql-example}, the root level object is \texttt{person}. The \texttt{person} object contains a field, \texttt{pet} which also is an object. From \texttt{pet} the \texttt{name} field is selected. In this example there were two related objects (\texttt{person} and \texttt{pet}) and one scalar field (\texttt{name}).

There are several \textit{scalar} field types defined in the GraphQL specification such as \textit{Int, Float, String, Boolean} and \textit{ID}, where \textit{ID} is a special type representing a unique identifier in the business domain. In addition to the scalar types, GraphQL also defines an enumeration type and a list type.

An object field can have arguments. These can be used to, for example, limit a query selection in a search. Imagine an object, \texttt{Project}, that contains the field \texttt{name}. For the client to be able to select from which project the name should be selected from, the schema can define an argument such as \texttt{path}. The client would then execute a query as in: \mint{js}|{project(path:"projects/project1"){name}}|

In addition to queries, a GraphQL schema can also define \textit{Mutations}.
\textit{Mutations} define types that perform mutating operations on the server. In the context of GraphQL, mutations refer to changing the state of a data-structure (for example changing the field value of \textit{name} of a \textit{Person} object). There is thus a separation between \textit{Mutation} and \textit{Query} types to make it clear to a client that performing a query will not change any resources on the server. This is however completely up to the server implementation to honor.

\subsection{Property-based Testing}
Property-based testing (PBT) was first introduced as a concept by Fink \cite{Fink97property-basedtesting} but today it is synonymous with the random testing introduced by Claessen et al.~in the tool QuickCheck \cite{ClaessenQuickCheck}. PBT is a testing technique that aims to verify the invariant properties of a system under test (SUT). Test cases are randomly produced and the PBT tools then try to find random values that prove the invariant wrong. If such an example is found it is a failing test case. Random values are produced by invoking \textit{generators}. A PBT library typically delivers generators for basic scalar types, such as integers, strings and booleans, and generators that are the combination of other generators. In this way, a user can combine generators to randomly produce relevant values for the domain under test.

Compared to example-based tests, PBT requires a shift in thinking. Instead of thinking of the specific output produced to a specific example input, we must instead think in invariants. For example, imagine that we test a search function. The search function requires a search string as input and returns a list of items that match the search string. Instead of using a predefined search string with a predefined outcome, we instead use a string generator for the input string and formulate the invariant properties of the resulting list of items. Such properties include (1) the name of the resulting entities should always contain the search string as a substring in their name (2) the returned entities should always conform to a given format and (3) the function should always return a successful result, i.e., not crash. With this generator and properties, we can start to generate any number of test cases to try and find cases where the properties are proven wrong.

In the implementation of our proposed method, we have leveraged a property-based approach and defined a generator to randomly produce GraphQL queries.

\begin{figure}
    \centering
\begin{minted}{json}
{"data": {
    "__schema": {
      "types": [{
          "kind": "OBJECT",
          "name": "Project",
          "description": null,
          "fields": [
            {
              "name": "description",
              "description": null,
              "args": [],
              "type": {
                "kind": "SCALAR",
                "name": "String",
                "ofType": null}},
            {
              "name": "id",
              "description": null,
              "args": [],
              "type": {
                "kind": "NON_NULL",
                "name": null,
                "ofType": {
                  "kind": "SCALAR",
                  "name": "ID",
                  "ofType": null}}},
            {
              "name": "members",
              "description": null,
              "args": [],
              "type": {
                "kind": "LIST",
                "name": null,
                "ofType": {
                  "kind": "OBJECT",
                  "name": "User",
                  "ofType": null}}},
            {
              "name": "name",
              "description": null,
              "args": [],
              "type": {
                "kind": "SCALAR",
                "name": "String",
                "ofType": null}},
            {
              "name": "owner",
              "description": null,
              "args": [],
              "type": {
                "kind": "OBJECT",
                "name": "User",
                "ofType": null}}]}]}}}
\end{minted}
    \caption{Project example schema type}
    \label{fig:schema-example}
\end{figure}

\begin{figure*}[h]
    \includegraphics[width=\textwidth]{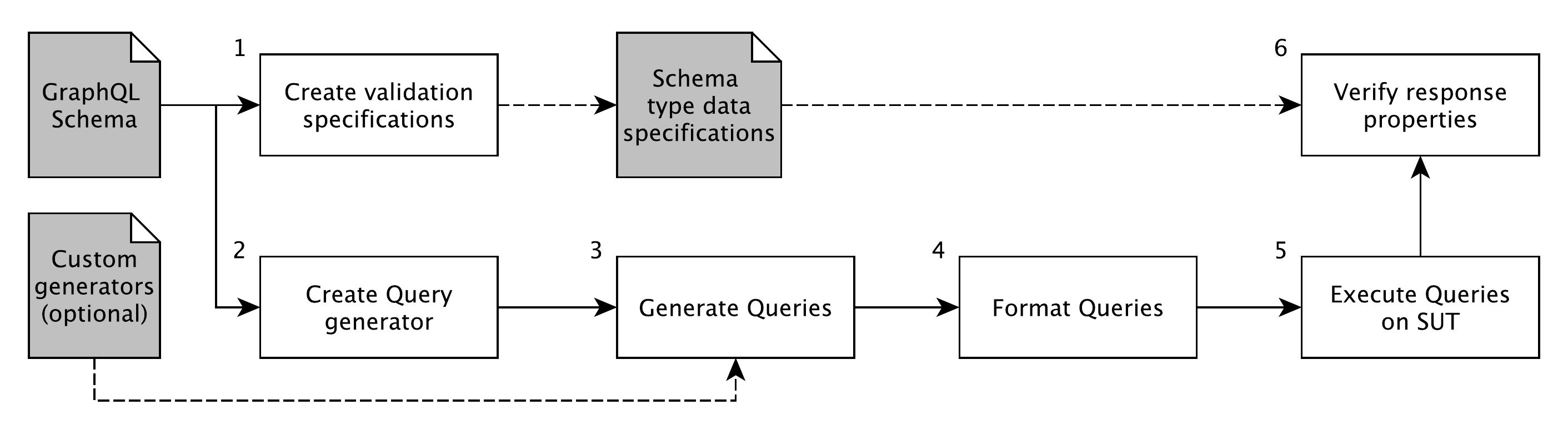}
    \caption{Overview of the test case generation method}
    \label{fig:method-overview}
\end{figure*}

\section{Schema Coverage}\label{section:schema-coverage}
In order to assess the effectiveness of a test suite, a coverage criterion is commonly used. However, the goal of a test suite is to ensure the quality of a software system and find any bugs present in the system. Thus, a measurement of coverage can be seen as a proxy for the fault finding capability of a test suite. In this paper, we evaluate the fault finding capability of our proposed method. Since our approach is based on random testing, we also have to ask ourselves, is the complete schema covered while tests are randomly generated? A deterministic method could be certain to generate tests for the full schema, given a sufficiently large test budget, but this is not as given with a method based on randomness. To evaluate the effectiveness of our approach in generating tests that cover the complete schema, we need to define a coverage criterion.   

For GraphQL APIs, observed in a black-box approach, we propose a schema coverage criterion. This coverage criterion is based on the \textit{objects} and their \textit{fields} defined for all the types in the schema. To be effective, generated test cases should cover all the available objects and fields defined in the schema.

We propose that each field defined in the schema is expressed in a tuple with its containing object. We use the tuple format to express all the unique pairs since field names can be the same on different objects. This method results in the set $\{[Object_t, Field_n]\}$ containing the tuples of all fields, $n$, of the objects of type $t$ in the schema. For example, imagine a \textit{Person} object with the available fields of \textit{name} and \textit{age} and a \textit{Pet} object with a field of \textit{name}. The resulting tuple set for the schema in this example would then be represented as $\{[Person, name][Person, age][Pet, name]\}$. During test generation, the same kind of tuples are created. A query of \texttt{\{person \{name age\}\}} would cover the tuples $\{[Person, name][Person, age]\}$. To measure which parts of the schema that have been tested by the generated queries, we observe how many of the possible tuples of the schema are present in the generated query data. Tuples in the schema which are not present in the generated test are thus not tested and not covered.

This coverage metric can be used both online and offline, meaning that the generated tests can be evaluated on how much of the schema is covered and once executed, the results from the system under test can be converted into tuples and assessed for coverage. In summary, since no established coverage metric exist for black-box GraphQL test evaluation, we propose to start with a basic and intuitive one.

\section{Proposed Method}

Our proposed method automatically generates tests based on a graph-based schema of possible operations. In our case, we have applied our method to GraphQL schema. As input, the method requires a GraphQL schema and, optionally, generators for the specific types in the schema. Providing such generators allows the user to control the generation of argument values. For example, the system under test might have a specific custom format to follow for certain values. A custom generator can then be supplied to create values that conform to the domain specification. In our case study, the SUT required a path-like format of arguments representing an \textit{id}, such as \texttt{"gid://gitlab/Issue/123"}. A string generator would likely never produce a random string matching this format. A user can then provide a custom generator that generates a random integer (for the "123" part), a random element from the set of \{"Issue", "Project", "Group"\} and combine those randomly generated parts into a valid format string. The result would be, \texttt{"gid://gitlab/<random element>/<random int>"}, a string with both static and random parts. This approach could be taken even further, by for example, query the SUT for valid issue numbers and only randomly select from those when generating ids. This would ensure that the generated tests use existing entities in the SUT. 

Tests are generated as a query to be executed on the system under test and an oracle, a way to assert the results of the test, that verify invariant properties of the system under test. The queries are randomly generated based on the available \textit{objects} and \textit{fields} in the schema. Since a schema \textit{object} can contain both \textit{scalar} and \textit{object} fields, the randomization of query fields is recursively resolved to be able to select fields available to the current object being processed. This means that starting with a root object we randomly select fields from that object. Any fields that are themselves an object will have nested fields that need to be resolved in the next recursion. This recursion continues until there are no more \textit{object} fields to resolve or a recursion limit is hit.

Since the method targets cyclic graph schemata the recursion depth is taken into consideration. We do this by allowing the user to control the number of fields selected for each depth of the query generation. This is explained in more detail in Section \ref{section-field-depth}.

Query fields can require input arguments, for example, to limit a selection by a search string. The schema expresses the types of these arguments and our test generation process uses generators for the specific types. The argument generators produce random values to be used as arguments in the query. A sizing parameter is used, as is common in property-based approaches, to control the size of generated arguments. This allows for our generators to start with the generation of simple arguments, such as an empty string, and continuously increase the complexity of the argument parameter value. The reasoning being to fail fast on arguments that are faster to produce.

Depending on the given parameters, the size of the generated queries will vary. This allows a user to start with the generation of smaller queries and, as the confidence in the API increases, increase the size of the queries. It should also be noted that in the case of GraphQL the SUT might have restrictions on the complexity of the queries. Such restrictions are put in place to protect the SUT from spending too many resources on executing the query. It is then necessary to be able to control the size of the queries to produce queries that pass the complexity validation and are executed.

Figure \ref{fig:method-overview} present an overview of the proposed method including the following steps:
\begin{enumerate}
    \item From the \textit{Schema}, generate specifications of the types in the \textit{Schema}. 
    \item From the \textit{Schema}, create a generator capable of randomly generating a list of query objects with fields.
    \item Generate \textit{n} number of queries.
    \item Given the generated query data, transform the query in GraphQL format.
    \item Execute the queries on the SUT.
    \item Evaluate the result in a property-based fashion
\end{enumerate}{}

\begin{figure}
    \centering
\begin{minted}{clj}
{:object/Query
 [:map
  [:fields
   [:map
    ["person"
     [:map
      [:type [:ref :object/Person]]]]]]]

 :object/Person
 [:map
  [:fields
   [:map
    ["pet" {:optional true} [:ref :object/Pet]]]]]

 :object/Pet
 [:map
  [:fields
   [:map
    ["name" [:gql.scalar/string]]]]]}
\end{minted}
    \caption{Illustrative example of specifications generated from a schema.}
    \label{fig:person-pet-spec-example}
\end{figure}

\subsection{Creation of type data specifications}

To verify that the returned result of executing a GraphQL query conforms to the expected result, we need specifications of the types and their relations from the given schema. As a one-time operation, we generate specifications that correspond to the objects and their fields in the schema. Figure \ref{fig:person-pet-spec-example} shows a specification for the 
schema example queried in Figure \ref{fig:person-gql-example}.

The \textit{Query} (1), describing the root level entities of a valid query, \textit{Person} (9) and \textit{Pet} (15) are all specified as \textit{Objects} since they are composites and contains \textit{fields}. The specification contains references for fields that are referring to another object as its type, such as the \textit{Person} object referencing a \textit{Pet} as the type of its field \textit{pet} (13). The \textit{Pet} object shows an example of a field, \textit{name}, being of a scalar type (19). This specification is then used to verify that returned data from a query, as in Figure \ref{fig:person-gql-example-payload}, conforms to the Schema of the SUT.

The input of this step is a GraphQL Schema and the output are specifications of all object types and their fields defined in the schema.

\subsection{Creation of the query generator}\label{section-field-depth}
To be able to generate GraphQL queries a generator is needed. Such a generator can be generic in that it can generate queries given any schema conforming to the GraphQL specification.

Queries in GraphQL can be recursive and cyclic. This must be taken into consideration when implementing a query generator. In our method, we propose to instead of making a depth recursive generator, make an iteration based generator. As already mentioned, a query schema contains objects with fields where fields can be scalar types or objects. A random generator must then first randomly generate the object to start with of the available query types, i.e., the root object. The fields to randomly generate must be selected from the available fields of the root object. This means that the generator must have the context of the current object available to make the field selection. We can then either do a depth-first recursion of expanding all fields that are objects, and for the new fields that are objects expand those further, etc., or we can make an iterative flat process where only the currently generated fields are expanded.

We found that having a schema with a low level of recursions in the types, a depth-first approach could be used. However, with a larger and more recursive schema it did not scale and the iterative approach described generate queries in a more controlled and timely fashion.

We created our generator to be instantiated with four parameters: (i) the collection of valid query roots, the types which can be the root of a query; (ii) the complete list of types for the schema, needed for lookup when generating; (iii) a number representing the maximum allowable fields to generate for an object in each iteration; and (iv) a function that provides generators for all the scalar types. 

\begin{figure}
    \centering
\begin{minted}{clj}
(repeatedly 10 #(gen/generate gen/string-alphanumeric 2))
=> ("4" "K3" "34" "8M" "HN" "6v" "u" "t2" "" "")

(repeatedly 5 #(gen/generate gen/string-alphanumeric 100))
=> 
("IB9T59O0G2vCVgQBXG61WXU143Jg9026OQ1bongGY01rml7s3YfVLmRgATyR4"
 "a6F1S9l8abT87Iz"
 "W82"
 "6EiJ3Md2OYizN"
 "")

(repeatedly 4 #(gen/generate gen/uuid 2))
=>
(#uuid "6f5322b3-25c7-4c1e-a6ab-bf411c6835ba"
 #uuid "b848d13c-640d-4a10-acf2-3a81e063fe3a"
 #uuid "b864ef51-1cc8-42ea-a3f3-d967872c4e4f"
 #uuid "1fe4a060-5fe5-4198-9960-2c1ae5649997")

(repeatedly 4 #(gen/generate gen/uuid 100))
=>
(#uuid "4903ce91-2125-47fb-86c4-9423a3755f4f"
 #uuid "2badf60b-a66b-48b7-b0f3-7b374e8cd166"
 #uuid "e38b58d7-4e51-42ff-8900-4c0fe0cf7ece"
 #uuid "9309ad7f-7092-4c08-ae97-c263bfdfdc05")
\end{minted}
    \caption{Example of the impact of size on generators}
    \label{fig:size-example}
\end{figure}

\begin{figure}
    \centering
\begin{minted}{clj}
[{:name "projects",
  :generation 0,
  :args
  [{:name "id",
    :type
    {:kind "NON_NULL",
     :name nil,
     :ofType {:kind "SCALAR", :name "ID", :ofType nil}},
     :value "7x8Z"}],
  :fields
  [{:name "description",
    :args [],
    :type {:kind "SCALAR", :name "String", :ofType nil}}
    {:name "members",
     :args [],
     :type
     {:kind "LIST",
      :name nil,
      :ofType {:kind "OBJECT", :name "User", :ofType nil}}}],
   :type {:kind "OBJECT", :name "Project", :ofType nil},
   :id #uuid "c02efeee-f161-42b4-b589-f2ca34692e8a"}
  {:name "description",
   :generation 1,
   :args [],
   :type {:kind "SCALAR", :name "String", :ofType nil},
   :field-id #uuid "c02efeee-f161-42b4-b589-f2ca34692e8a",
   :from-type "Project"}
  {:name "members",
   :generation 1,
   :args [],
   :fields
   [{:name "projects",
     :args [],
     :type
     {:kind "LIST",
      :name nil,
      :ofType {:kind "OBJECT", :name "Project", :ofType nil}}}
    {:name "name",
     :args [],
     :type {:kind "SCALAR", :name "string", :ofType nil}}],
   :type
   {:kind "LIST",
    :name nil,
    :ofType {:kind "OBJECT", :name "User", :ofType nil}},
   :id #uuid "9c358b36-b5f8-42c6-8b80-aef3cac6f5f1"
   :field-id #uuid "c02efeee-f161-42b4-b589-f2ca34692e8a",
   :from-type "Project"},
  {:name "name",
   :args [],
   :type {:kind "SCALAR", :name "String", :ofType nil},
   :field-id #uuid "9c358b36-b5f8-42c6-8b80-aef3cac6f5f1",
   :from-type "User"}
  {:name "projects",
   :args [],
   :type
   {:kind "LIST",
    :name nil,
    :ofType {:kind "OBJECT", :name "Project", :ofType nil}},
   :field-id #uuid "9c358b36-b5f8-42c6-8b80-aef3cac6f5f1",
   :from-type "User"}]
\end{minted}
    \caption{Randomly generated schema nodes}
    \label{fig:generated-nodes-example}
\end{figure}

\subsection{Query generation}
Figure \ref{fig:generated-nodes-example} shows an example of the flat list of query nodes produced by the generator. The root of the query is the \texttt{"projects"} in line 1. This object has a generated argument, (3-9), with the generated value of "7x82" (9). There are two fields randomly generated, \texttt{"description"} (11) and \texttt{"members"} (14). For each generated node which is an object we attach a unique id (line 21 for \texttt{project}) and for each generated field we attach a \textit{field-id} which has the same value as the object the field belongs to. By doing this we can later create a tree from the flat list of nodes.

During generation, the flat list is iterated and each object or field which is processed gets a \texttt{generation} value attached. The generation value is used to keep track of which nodes have already been processed and which nodes that need processing in the current iteration.

To restrict the number of nodes generated, the depth of the resulting query, a sizing parameter is used. It is common for property-based random generators to have a size parameter, without one we have no control of the different sizes of random values produced by the generator. As an example, consider the \texttt{gen/string} generator and a generator for a universally unique identifier\footnote{\url{https://en.wikipedia.org/wiki/Universally_unique_identifier}} (UUID), \texttt{gen/uuid}. In the case of the \texttt{gen/string}, the size parameter will be used to produce random strings with a length proportional to the size parameter. Figure \ref{fig:size-example} shows generated strings from a size parameter value of 2, in line 1-2, and a size parameter value of 100, in line 4-10. As we can see there is a significant difference in the length of the strings generated. On the other hand, the \texttt{gen/uuid} generator show no difference in the size of the output in Figure \ref{fig:size-example} lines 12-17, for size 2, and lines 19-24, for size 100. The reason for this is that a UUID uses a fixed 128-bit format, generating anything shorter or larger does not make sense, thus the \texttt{gen/uuid} disregards the size parameter. 

When a PBT library generates test cases, it will start with a small size parameter value and increase it. The reason is to first try to find faults with simpler values before generating more complex values, as a simple example is preferred for reproducibility of a failure. Our method uses the size parameter in two ways. First, for any argument generation, such as the argument value in Figure \ref{fig:generated-nodes-example} line 9 ("7x82"). This means that any argument value we produce will start with small values and increase the complexity as the random testing increases the size parameter. Secondly, as mentioned, we use it to restrict the numbers of query nodes generated in the flat list. This is done by controlling the number of generating iterations of query nodes to be the minimum of the current \textit{size} and the \textit{maximum-allowed-iterations}. This means that our query generator starts with the generation of a few number of nodes and increases the number of nodes as the size parameter increases, but can be capped at a \texttt{maximum-allowed} value to restrict the depth.
Expansion of new object fields will stop at this generation size. However, in many cases the depth will be randomly controlled by the number of fields generated for objects, if only scalar fields are generated for the current generation, there is no more field generation to perform for the next generation.

\begin{figure}
    \centering
    \includegraphics[width=\linewidth]{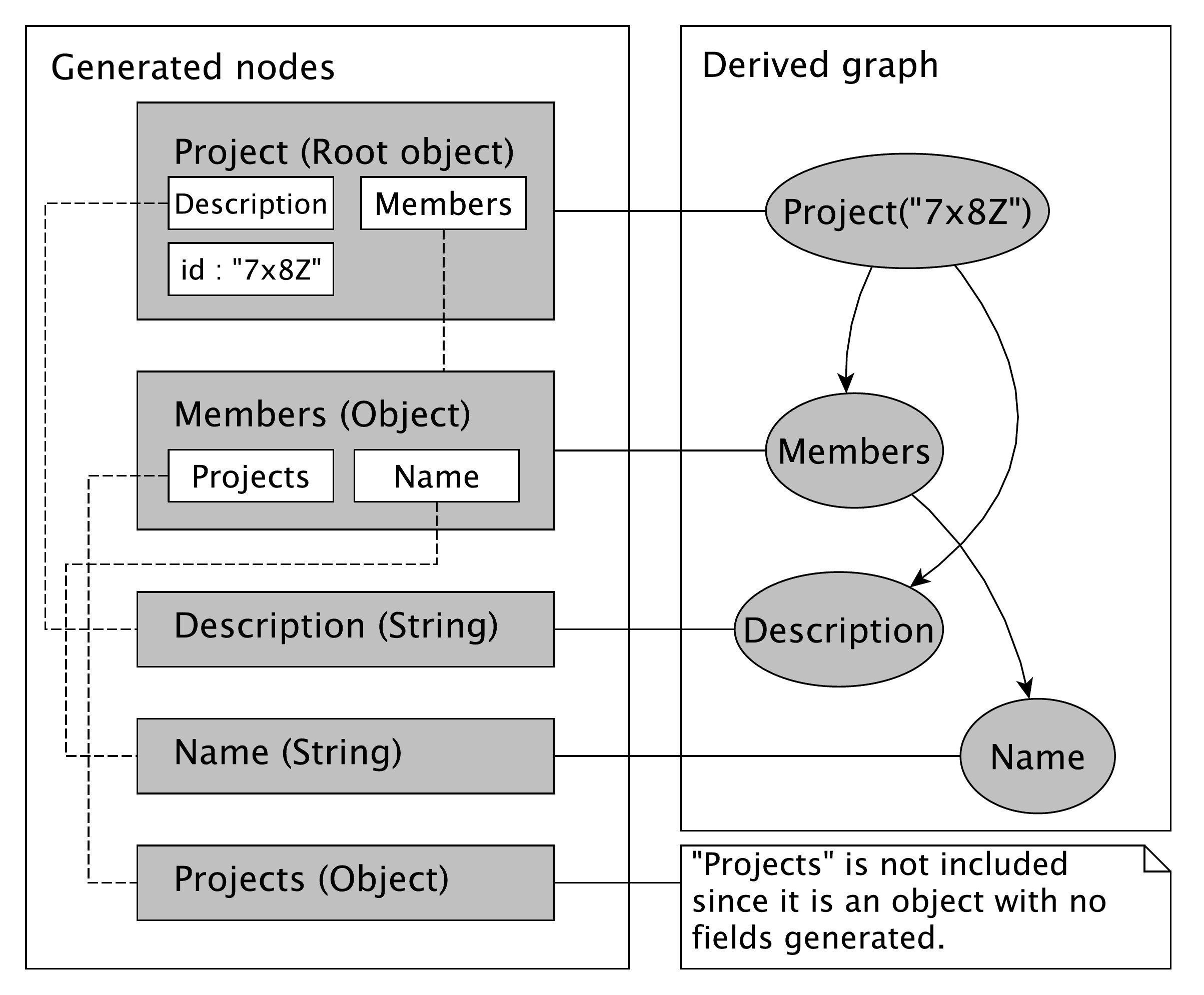}
    \caption{Relationships of generated nodes to the query graph. Dashed lines are the object-id to field-id relation. Solid lines are the relation of the nodes to the query. Arrows are the relations in the query graph.}
    \label{fig:generation}
\end{figure}

\subsection{Query creation}
In this step of the method, we first use the flat list of randomly generated query nodes and transforms them into a tree. This is done by using the object id to field-id connection produced in the query generation. Figure \ref{fig:generation} shows a graphical representation of the relationships between the flat list of nodes and the query graph.
This process starts from the last generation down to the first, i.e., the last fields to be generated are the bottom leafs of the tree and those should first be inserted into their containing object as the containing object can be a leaf to another object. By this method, we end up with a tree as in Figure \ref{fig:tree-example}. In Figure \ref{fig:tree-example} we can see the flat list from Figure \ref{fig:generated-nodes-example} transformed into its tree form.

In this step, some cleaning of the flat list is also done before the tree is created. This process is simplified by using the flat list approach, which we propose, instead of letting the generator generate trees. An example of a cleaning operation is to remove any fields from the last generation with the type of \textit{object}. To be a valid query, an object must have a field selection but the generation limit was hit before these were processed and they should be removed from the resulting tree.

With the resulting tree structure, we make the final transformation of turning the tree data structure into an actual executable GraphQL query. This transformation recursively processes the tree and composes the name of the fields with the required delimiters in the resulting string. Figure \ref{fig:random-gql-example} shows the final executable output of the tree data structure of the query in Figure \ref{fig:tree-example}. It is notable to mention, this method creates test cases that are human-readable. The resulting query can be executed manually or stored as input to any resulting bug report. 

\subsection{Execution and result verification}
As this is a property-based method we need both a generator for our test data and properties to verify the results of executing the test data. We have created the query generator as previously described. Running test cases are done by letting the generator generate a query, execute the query on the system under test and verify the stated properties of the result. Given the generator, the properties and a means of executing the test cases, generation can be handled by a PBT library.

We suggest two properties, which were also used in our evaluation. The first one is to verify the HTTP status code returned by executing the query. HTTP has well-defined status codes. For example, a "200" status code is the "OK" status code, a success, while "500" means "Internal Server Error". Thus a status code other than "200" would result in a failed test. The second suggested property is to verify that the resulting data returned conforms to the given schema, i.e, if the query is for a \textit{Project} with a field of \textit{name}, this property verify that the response returned include only valid fields for a project, name, and that the name scalar contains a value of the correct scalar type, a string.

Being a property-based method a user can always add more properties that better verify the specific domain properties of the system under test.

\begin{figure}
    \centering
\begin{minted}{clj}
[{:name "projects",
  :type {:kind "OBJECT", :name "Project", :ofType nil},
  :id #uuid "c02efeee-f161-42b4-b589-f2ca34692e8a",
  :generation 0,
  :args
  [{:name "id",
    :type
    {:kind "NON_NULL",
     :name nil,
     :ofType {:kind "SCALAR", :name "ID", :ofType nil}},
    :value "7x8Z"}],
  :fields
  [{:generation 1,
    :name "description",
    :args [],
    :fields [],
    :type {:kind "SCALAR", :name "String", :ofType nil},
    :field-id #uuid "c02efeee-f161-42b4-b589-f2ca34692e8a",
    :from-type "Project"}
   {:name "members",
    :generation 1,
    :args [],
    :fields
    [{:name "name",
      :args [],
      :type {:kind "SCALAR", :name "String", :ofType nil},
      :field-id #uuid "9c358b36-b5f8-42c6-8b80-aef3cac6f5f1",
      :from-type "User"}],
    :type
    {:kind "LIST",
     :name nil,
     :ofType {:kind "OBJECT", :name "User", :ofType nil}},
    :id #uuid "9c358b36-b5f8-42c6-8b80-aef3cac6f5f1",
    :field-id #uuid "c02efeee-f161-42b4-b589-f2ca34692e8a",
    :from-type "Project"}]}]
\end{minted}
    \caption{Composed tree of randomly generated schema nodes}
    \label{fig:tree-example}
\end{figure}

\begin{figure}
    \centering
\begin{minted}{text}
{projects(id: "7x8Z"){description members{name}}}
\end{minted}
    \caption{Randomly generated example in GraphQL}
    \label{fig:random-gql-example}
\end{figure}

\section{Evaluation}

In this section, we provide an evaluation of the proof-of-concept (PoC) method implementation applied to a real world software system and a controlled example application, with the goal of evaluating the schema coverage and fault finding capability. The PoC was implemented in the Clojure programming language\cite{hickey:AHistoryOfClojure} and is available as a replication package\footnote{\url{https://github.com/zclj/replication-packages/tree/master/GraphQL}}. We used the PBT library TestCheck\footnote{https://github.com/clojure/test.check} to build generators, and to run the test cases. Basic generators from TestCheck was used as primitives to create the more complex GraphQL specific generators based on the schema. We implemented a black-box version of the method, meaning we did not assume any access to the source code or any other internal artifact of the SUT. The only input required was the GraphQL Schema. Since some types are only used in the Query part or in the Mutation part of the schema, the possibility to manually filter fields from inclusion in coverage evaluation were made available, in the case where it is desired to only do coverage metrics on parts on the schema. Such a case would be if only a subset of the schema is tested.

We do not provide any comparison with tools targeting REST APIs. While both REST and GraphQL exposes Web APIs, they are very different methods and thus any tool specific to one is not applicable to the other.

For the purpose of this evaluation, we formulated the following research questions:
\begin{itemize}
    \item \textbf{RQ1}: What schema coverage can be achieved by the method?
    \item \textbf{RQ2}: What is the fault finding capability of the method?
\end{itemize}

The focus of the first research question is to measure if the method can cover a complete schema and the effect the inputs to the method have on the coverage of the schema. We additionally set out to evaluate the efficiency of generating achievable coverage.

The second research question evaluates if the method can find both real faults and injected faults by generating random queries and input arguments executed on the system under test.

\subsection{Studied Cases}

We evaluated our method using two different cases. One example application where we could control the specific errors injected and one real industry system, GitLab.

\subsubsection{Example application}
We implemented a small example application with a small GraphQL Schema, shown in Figure \ref{fig:schema-example}. The purpose of the application was as a SUT with known faults, to allow for evaluation of the fault finding capability of the method.

\subsubsection{GitLab}
For our evaluation of an industry system we have used GitLab as the system under test. GitLab is a product that provides tools for the complete DevOps lifecycle. This includes tools such as code version control, issue handling and build-pipelines. GitLab have been recognised as a leader in cloud-native continuous integration by Forrester\footnote{https://go.forrester.com} \cite{GitLabLeaderInCi}.

GitLab is provided both as a Software-As-A-Service offering and an on-premise product. The product also comes in two different versions; the Enterprise Edition (EE) and the Community Edition (CE) which differs in features and support options. GitLab CE is available as open-source software, has a GraphQL API, is a real industry software with a large usage, and with the combination of being able to run on-premise, it is a good industry case study for our method.

Many services are using GraphQL APIs. However, to be able to run thousands of test cases, a hosted service is not a viable option. Industry grade services use rate limits and denial-of-service protections to not allow clients to abuse the service. Therefore we need to run our experiments in a local setup with software that is available to do so.

\subsection{Experimental Setup and Method}

We ran GitLab with the officially provided Docker\footnote{https://www.docker.com/} container\footnote{https://docs.gitlab.com/omnibus/docker/}. The coverage and fault finding experiments were executed on commodity developer hardware, a MacBook Pro, 2,9 GHz i9, 16 GB RAM.

\subsubsection{RQ1: schema coverage} 

To measure the schema coverage of the generated queries we created object-field tuples of both the GitLab GraphQL schema and the generated query data. We used the tuple format described in Section \ref{section:schema-coverage} to produce all the unique pairs. The goal of the generated query data is to cover all such tuples and in doing so get coverage on all available fields in the schema.

We generated \textit{n} queries with a \textit{max-field} limit of \textit{m}. The field limit controls the max number of fields to be generated for each object. A low number will decrease the likelihood of a deep query and a high number will increase the likelihood of a deep query.

Each set of parameter values, \textit{n} and \textit{m} was executed 30 times. The average coverage was reported but also the aggregated coverage over all 30 iterations. The reason for measuring the aggregated coverage is due to the possible complexity limits of the SUT. Very deep queries have high coverage, but the query might not be allowed to execute on the SUT. The aggregated coverage then allows evaluating if high coverage is achieved with smaller queries but with a higher number of executions.

\textit{RQ2: Fault finding capability} To evaluate the fault finding capabilities of our method, we applied it to GitLab. To evaluate fault finding, queries must be executed on a running service, a local GitLab instance in our case.

The full process in Figure \ref{fig:method-overview} was followed. Queries were generated from the schema, formatted and executed on the SUT and properties of the results returned were verified. A property was defined to check for the presence of "Internal server errors", i.e, status code 500, in the returned result. Any such returned result is the effect of an unhandled crash on the server and is categorized as a fault. In addition, a property verifying that the result payload conforms to its specific schema type was also defined.

In addition to the GitLab evaluation, we sought to evaluate which type of faults the method can find. For this purpose, we used an example application with seeded faults. The GraphQL schema of the application contained the types \texttt{Project} (with fields id, name, description, owner, members) and \texttt{User} (with fields id, name, age, projects). The \texttt{owner} and \texttt{members} of a project are references to entities of the \texttt{User} type and the \texttt{projects} field of a \texttt{User} is a reference to a \texttt{Project}, making the schema cyclic.

Each object reference in the GraphQL query will be handled by a \textit{resolver}. A resolver contain the logic of finding the data for that query node. For example, a resolver for \texttt{Project} with an \texttt{Id} argument could connect to a database and select the project with the given id, fulfilling that part of the query. To handle the example schema we needed four resolvers. The logic for those resolvers were the target for the injected bugs. The injected bugs were reviewed by senior industrial developers and deemed as plausible.

In total 15 bugs were injected.
\begin{itemize}
    \item 3 input validation bugs, for the project resolver.
    \item 4 logic bugs, one for each resolver
    \item 4 bugs where filtering is done using the wrong field, i.e., trying to find an entity by name instead of by id when given the value for the id. One for each resolver.
    \item 4 bugs returning the wrong type, i.e., returning a list were a single value is expected. One for each resolver.
\end{itemize}{}

We executed the generated test cases on our working implementation without finding any problems, it was also reviewed with no known problems. We then injected each of the described bugs, one at a time. With a planted bug in place, test cases were then generated and executed as with the setup previously described.

\subsection{Results}

In this section we provide the results of the case studies performed on GitLab and our example application.

\subsubsection{RQ1: Schema coverage}

It turns out that our method can produce queries with full coverage of the fields defined in the schema. As can be seen in Table \ref{tab:coverage-result}, a high coverage can be achieved both by running shallow queries a large number of times or deeper queries a fewer number of times. However, before deciding on which set of parameters to use, the SUT must be considered. If the complexity limits of the SUT will not allow deep queries a larger number of shallow queries can be generated for the same achieved coverage. 
In addition, deep queries will carry a larger cost in processing. It will be more resource-intensive to generate a GraphQL query string, serialize and send to the server for deserialization than with smaller queries.

With the results from Table \ref{tab:coverage-result} we can further examine at which iteration count the full coverage was achieved. Each configuration of the parameter values was executed until 100\% coverage was achieved for 30 times. The average iteration count required for 100\% coverage is shown in Table \ref{tab:iterations-result}. This result, in combination with Table \ref{tab:coverage-result}, gives more information about the trade-off between size, field selection size, execution time and coverage to consider for the specific SUT.

In summary, our method can achieve full coverage of the schema under test with different configurations accommodating the complexity properties of the system under test.

\begin{result}
Full schema coverage can be achieved by randomly generating test cases, both for many smaller queries or with fewer larger ones. However, there is a trade-off between query size, complexity properties of the system under test and efficiency.
\end{result}

\begin{table}[h]
    \centering
    \caption{Schema coverage result on GitLab schema}
    \begin{tabular}{l|c|l|l|l}
        \hline
        size & max-limit & avg. & total & time (s)\\
        \hline
        10 & 1 & 3.18\% & 20.58\% & 0.03 \\
        100 & 1 & 14.04\% & 49.59\% & 0.62 \\
        1000 & 1 & 36.95\% & 87.24\% & 9.12 \\
        10000 & 1 & 73.33\% & 100.00\% & 94.83  \\
        \hline
        10 & 2 & 6.56\% & 38.89\% & 0.04 \\
        100 & 2 & 28.94\% & 85.60\% & 1.26 \\
        1000 & 2 & 64.87\% & 100.00\% & 19.32 \\
        10000 & 2 & 97.96\% & 100.00\% & 199.65 \\
        \hline
        10 & 3 & 12.00\% & 54.94\% & 0.07 \\
        100 & 3 & 49.08\% & 99.59\% & 2.56 \\
        1000 & 3 & 90.62\% & 100.00\% & 40.97 \\
        10000 & 3 & 100.00\% & 100.00\% & 416.31  \\
        \hline
        10 & 4 & 20.86\% & 77.16\% & 0.11 \\
        100 & 4 & 71.51\% & 99.79\% & 5.73 \\
        1000 & 4 & 99.29\% & 100.00\% & 93.57 \\
        10000 & 4 & 100.00\% & 100.00\% & 931.63  \\
        \hline
        \hline
    \end{tabular}
    
    \label{tab:coverage-result}
\end{table}{}

\begin{table}[h]
    \centering
    \caption{Number of iterations, on average, required for 100\% schema coverage result on GitLab schema}
    \begin{tabular}{l|c|c|c|c}
        \hline
         size &  max-limit 1 & max-limit 2 & max-limit 3 & max-limit 4 \\
         \hline
         100 & - & - & 61.43 & 18.0 \\
         1000 & - & 26.63 & 6.57 & 2.2 \\
         10000 & 26.0 & 3.77 & 1.13 & 1.0 \\ 
         \hline
        \hline
    \end{tabular}
    
    \label{tab:iterations-result}
\end{table}{}

\subsubsection{RQ2: Fault finding capability}

Regarding the evaluation of the example application with seeded faults, in total, 11 of the 15 bugs were found (73\%). Bugs were found in all of the four resolvers of the domain entities, this means that the generated test queries covered all resolvers. However, to accomplish this, custom logic was needed in the provided generators. For example, a \texttt{Project} object contains a reference to its owner, which is a \texttt{User} object. The \texttt{Project} object takes an \textit{id} as an argument, which is used to find the specific project. To be able to reach the \texttt{User} part of the \texttt{Project} we need an id of an existing project. Otherwise, the query result will be empty for the non-existing project id. This means that a user-provided custom generator that generates valid ids, for the SUT, is needed to get execution of the nested resolvers.

All 3 input validation bugs were found as well as the 4 logic bugs without any specific considerations other than those described above. The 4 bugs where the wrong type of data was returned for a given field used another property, to verify the result of executed queries, in addition to the property failing any "Internal Server Error" status codes. Queries that do not cause a crash, but for some reason is processed with errors, are returned to the client with a "Success" status code. In addition to the status code, the result will also include an "Error" section. The type error bugs we injected caused such errors. We then created a property that fails when an error is returned, even though the status code is a success, which found all the 4 type error bugs.

The 4 bugs not found where all the bugs were we filtered the entities requested with the wrong field, such as selecting for a project with \textit{name} "1" when \textit{id} was the correct field to select for. The effect for the client, in this case, is that instead of getting the project with \textit{id} "1", the client gets an empty result, since no project was found. Our properties did not "know" to expect and verify a non-empty result for the specific project id and thus did not fail the generated test. 

\begin{result}
All bugs resulting in a crash or an error message was found by our generated test cases. Bugs that result in the wrong response returned would need stronger domain aware properties to be found.
\end{result}

We found several, to the best of our knowledge, new unreported input validation issues in GitLab by applying our method, both for Query and Mutation types. The issues were reported to the project's issue tracking system with queries generated by our implementation as reproducible examples \cite{gitlabbug-1, gitlabbug-2, gitlabbug-3, gitlabbug-4, gitlabbug-5, gitlabbug-6, gitlabbug-7, gitlabbug-8}. Several of the issues have had developer attention and are scheduled for correction and inclusion in the product.
These issues were caused by inserting our generated strings as input to the query arguments which required input values. For the Query type, when our string generators produced alpha-numeric strings no faults were found. However, when string generators producing any char value, 0-255, were used, input validation bugs were detected. While good at finding bugs, the string generators produced a lot of invalid input. About 40\% of the executed queries returned with a client error of bad input. This indicate that the string generator produces input that finds bugs but at the same time produce a lot of queries that are not valid for further processing by the SUT.

One of the issues required a domain specific modification to the generated values. The bug only manifested when a specific query field was selected for an existing GitLab project. An existing project is referenced with an argument in the format \texttt{"<user>/<project>"}. When our generator was configured to return an existing project, the query to find the bug was produced. This shows that knowledge of domain entities need to be supplied to get a stronger test suite, without existing project entities and a way to generate their paths, we would not have been able to find the bug.

These results indicate that our proposed method, of random generation of queries and input values, can find real faults in industry-developed applications.

\begin{result}
Test cases in the form of randomly generated GraphQL queries can find real new faults in real industry software systems. There is a trade-off between the fault finding capability of different value generators and input deemed as valid by the system under test. To be more effective, custom generators can be used.
\end{result}{}

\subsubsection{Threats to Validity}

All findings presented are based on data gathered using our generators. The implementation of the GraphQL query generator is thus a threat to the internal validity of the results. While the exact numbers might vary with a different implementation, we have no reason to believe that another implementation following the proposed method would not reach the same conclusions as we have. 

Our proposed method have a large component of randomness based on the usage of random generators. Given this, the exact results very between executions. To mitigate such differences, we ran each configuration 30 times.

Considering the external validity, we applied our method on several publicly available APIs \cite{apiguru}. However APIs not requiring authentication or registrations are typical demonstrations or non-production APIs. With that in mind we focused our case-study, as reported, on an industry-grade API and a controlled example. The learnings from a set of public APIs was integrated in the solution used in our case-study. Those where (1) it is common for the schema to contain some domain specific information which is expressed in a scalar type (such as a string) but requires a specific format found in documentation. (2) For schema with a less recursive structure a more straight forward generation approach can be used, but with a recursive schema, control of the recursion size is needed to control generation. These two points was then integrated in the solution used on our real-world industry grade case-study, GitLab, by allowing for custom generators and to control the depth of the generation as described. 

With that said, the main evaluation was performed on one industrial case study and one smaller example software. Performing an evaluation on a larger set of industry-grade software systems would increase the confidence in the generalizability of the proposed method. However, since GraphQL has a defined specification, the method is applicable to different APIs than those used in the case-study.

\section{Related Work}

Since we are testing GraphQL APIs and using a PBT technique we present prior work in those areas and how it relates to ours. Also, since we are performing black-box testing of an API based on HTTP, like REST, black-box automatic REST API testing is of relevance, but not directly comparable due to the differences between GraphQL and REST. 

\subsection{GraphQL}
The only current work in GraphQL API testing, to the best of our knowledge, is a proposed method by Vargas et al. where a user provided initial test case is mutated by a set of "deviations" to produce test cases \cite{vargas-deviation-testing-2018}. This method differs from ours since we aim for a fully automatic method, where no user provided test cases are required. Also, we use a PBT approach where Vargas et al. use a mutation based approach.

The current state-of-the-practice in GraphQL testing is a manual practice, either by using tools that require manually created test-cases or recordings of manually created traffic, or manually coded tests. Since GraphQL uses the HTTP protocol, any testing tool with manually formulated test cases that target REST can be used, such as Postman.\footnote{https://www.postman.com/} However, a recent development is an automatic tool by Meeshkan,\footnote{\url{https://meeshkan.com/}} but its impact in practice remains to be seen.

\subsection{Automatic Black-box Web API Testing}
Several black-box methods to automatically generate tests for Web APIs have been proposed in recent years \cite{RESTTESTGEN, Atlidakis:2019:RSR:3339505.3339600, karlsson2019quickrest, ed-douibi:RESTAPIS}. These methods only require an artifact describing the API end-points, including the type of data required and the expected responses. The approaches are similar to ours since they only require a SUT and a specification to automatically generate tests of HTTP based web services. However, these methods target RESTful APIs described with the OpenAPI specification\footnote{https://www.openapis.org/} and not GraphQL as is our case. This means that those tools have no support to generate test cases from a GraphQL schema.

\subsection{Property-based Testing}

Property-based testing is a testing technique that have been applied to test systems in many different industry domains \cite{ArtsTelecom, HughesTestingTheHardStuff, HughesDropBox, ArtsAutosar}, including web services \cite{karlsson2019quickrest, Francisco:2013:TWS:2505305.2505306, PropErWebServiceTesting, Li:2014:APT:2543728.2543741, Fredlund:2014:PTJ:2680842.2681200}, with different levels of automation. However, the targets for web-based APIs have been for RESTful services, WSDL based services, and JSON Schema, no method for GraphQL based APIs have been proposed.

The most recent example, QuickREST, is an automatic method that applies property-based testing to RESTful APIs \cite{karlsson2019quickrest}. The technique used is similar to ours, an automatic black-box approach leveraging property-based testing, but QuickREST targets RESTful APIs and does not provide solutions for the specific challenges posed by testing GraphQL APIs. An example of such a challenge is the cyclic nature of a graph-based API where the randomly generated recursion depth must be controlled, which we do in this work.

\section{Discussion and Future Work}

We evaluated our method on GitLab. GitLab has parts of their complete API ported to GraphQL while the complete set of operations are still available in their REST API. This is an example of how one service API, GraphQL in this case, depends on resources created with another service, the REST API. Thus, for future work, it would be interesting to find an automatic method that can combine the test generation of both GraphQL and REST to be able to create, query and verify resources of a system of combined services. Industry systems that are in the transition from REST APIs to GraphQL APIs could also benefit from such a method, since then feature parity test cases could be generated between the functionality exposed by the REST API and GraphQL API.

To increase the fault finding effectiveness an automatic test generation method would benefit from more domain knowledge. However, it is preferable to not break automation and require a human to codify all knowledge. Thus it would be interesting to automatically find domain knowledge from running a set of generated test cases and improve future generation with the knowledge gained from previous executions. 

The basic schema coverage criteria proposed could be expanded to include the coverage of sub-graphs in the query. Such an extension would allow a user to define important paths to make sure such paths are covered in the generated test cases.

With this paper, we would like to highlight the importance of performing research in automatic testing of GraphQL APIs since its usage is increasing in industry and encourage other researchers to propose white-box methods to solve this task, complementary to our black-box approach. In doing so, practitioners would have a stronger set of proven methods to apply to test their GraphQL APIs.

\section{Conclusion}

In only a few years GraphQL has become a widely adopted technique to expose Web APIs in industry. However, currently, the testing of such APIs is a manual task, requiring practitioners to manually create test cases. With this work, we want to alleviate this pain point and make automatic test case generation of GraphQL APIs possible and effective.

In this paper, we propose, to the best of our knowledge, the first method to fully automatically generate test cases for GraphQL APIs, as well as a method of how to evaluate the coverage of such test cases. We have proposed a property-based method were generators are used to randomly produce test cases in the form of GraphQL queries and values for any arguments in the query.

Our evaluation results show how this method achieves high coverage, up the 100\% for many configurations, the sizing considerations to make when generating queries and the fault finding capabilities of the method. In addition to being an effective method in reaching a high coverage of the schema, the method is as well effective in finding real faults, as shown with the reported issues to GitLab. Further, given a different set of input parameters and custom generators, the method is capable of producing queries that are valid depending on the complexity requirements of the system under test and to produce domain-specific input values.

\section*{Acknowledgements}

This work is supported by ABB and the industrial postgraduate school Automation Region Research Academy (ARRAY) funded by The Knowledge Foundation.

\bibliographystyle{IEEEtran}
\bibliography{main}
\end{document}